\documentclass[preprint,11pt,3p]{elsarticle}

\usepackage{graphicx}
\usepackage{amssymb}

\usepackage[utf8]{inputenc}
\usepackage[T1]{fontenc}
\usepackage{float}
\usepackage[prependcaption, textsize=tiny, linecolor=red, backgroundcolor=red!25, bordercolor=red, textwidth=3.1cm]{todonotes}
\usepackage[plainpages=false, bookmarks=false, hidelinks]{hyperref}        

\hypersetup{
  pdftitle={Global Fire Season Severity Analysis and Forecasting}, 
  pdfauthor={Leonardo N Ferreira},
  pdfcreator={Leonardo N Ferreira} 
  colorlinks=true,          
  linkcolor=blue,           
  citecolor=blue,           
  filecolor=magenta,          
  urlcolor=blue,
  bookmarksdepth=2
}

\makeatletter
\def\ps@pprintTitle{%
   \let\@oddhead\@empty
   \let\@evenhead\@empty
   \let\@oddfoot\@empty
   \let\@evenfoot\@oddfoot
}
\makeatother

\begin{document}

\begin{frontmatter}


\title{Global Fire Season Severity Analysis and Forecasting\tnoteref{mytitlenote}}

\tnotetext[mytitlenote]{L.N.F conceived and conducted the experiments. All the authors analyzed the results, wrote and reviewed the paper. Declarations of interest: none}



\author[address1]{Leonardo N. Ferreira\corref{mycorrespondingauthor}}
\ead{leonardo.ferreira@inpe.br}
\cortext[mycorrespondingauthor]{Corresponding author}

\author[address2,address3]{Didier A. Vega-Oliveros}
\ead{davo@icmc.usp.br}

\author[address3]{Liang Zhao}
\ead{zhao@usp.br}

\author[address4]{\\Manoel F. Cardoso}
\ead{manoel.cardoso@inpe.br}

\author[address1,address5]{Elbert E. N. Macau}
\ead{elbert.macau@unifesp.br}

\address[address1]{Associated Laboratory for Computing and Applied Mathematics, National Institute for Space Research,\\S\~ao Jos\'{e} Dos Campos - SP, Brazil.}

\address[address2]{School of Informatics, Computing and Engineering, Indiana University, Bloomington, IN, USA}

\address[address3]{Department of Computing and Mathematics, University of S\~ao Paulo, Ribeir\~ao Preto - SP, Brazil.}

\address[address4]{Center for Earth System Science, National Institute for Space Research, Cachoeira Paulista - SP, Brazil.}

\address[address5]{Institute of Science and Technology, Federal University of S\~ao Paulo, S\~ao Jos\'{e} Dos Campos - SP, Brazil.}

\begin{abstract}

Fire activity has a huge impact on human lives. Different models have been proposed to predict fire activity, which can be classified into global and regional ones. Global fire models focus on longer timescale simulations and can be very complex. Regional fire models concentrate on seasonal forecasting but usually require inputs that are not available in many places. Motivated by the possibility of having a simple, fast, and general model, we propose a seasonal fire prediction methodology based on time series forecasting methods. It consists of dividing the studied area into grid cells and extracting time series of fire counts to fit the forecasting models. We apply these models to estimate the fire season severity (FSS) from each cell, here defined as the sum of the fire counts detected in a season. Experimental results using a global fire detection data set show that the proposed approach can predict FSS with a relatively low error in many regions. The proposed approach is reasonably fast and can be applied on a global scale. 
\end{abstract}

\begin{keyword}
Global fire activity \sep Wildfire \sep Fire season length \sep Fire severity \sep Climate change \sep Time series prediction


\end{keyword}

\end{frontmatter}


\section{Introduction}
\label{sec_introction}

Wildfires have a fundamental role in the environment and a huge impact on human lives \citep{pereira08,chen2010seasonal,Brando2014}. They affect biodiversity, cause damage to forests and properties, influence global climate, and pose direct risks to human health \citep{flannigan06,spessa15,dey18}. Forecasting fire activity thus brings many benefits for fire management and control \citep{pereira08,flannigan13}. 

Models have been used to predict fire occurrence in global and regional scales. Global fire models focus on long-term predictions (decades to centuries) and include a complex set of dynamic factors related to climate, vegetation, and human activity \citep{bowman09}, which are linked to the presence and flammability of vegetation biomass, and sources of ignitions \citep{arroyo08}. Their high level of detail and precision lead to satisfactory results in long-term predictions but were not intended for seasonal forecasting \citep{rabin17}. Regional models are mainly applied for projections at the seasonal timescale. They provide good forecasting results for specific regions but require input data like vegetation (fuel), topography and weather, that may not be available for other regions \citep{roads05}. This limitation makes their application difficult to specific regions \citep{Westerling2003,Roads2010ncep,chen11,Marcos2015,spessa15}. Some methods have been developed to predict seasonal fire activity on a global scale \citep{chen16,turco18}. In general, they use burned area data, which do not accurately account for smaller real fires \citep{hantson13,fornacca17,earl17,fornacca17}. All these mentioned disadvantages motivate the development of a simple, accurate, and widely-applicable seasonal forecasting method. 

In this paper, we propose a seasonal fire prediction approach, which is simple, effective, relatively fast, and can be generalized and used on a global scale. Our methodology is based on applying seven different time series forecasting methods and evaluating their suitability as seasonal fire forecasting predictors. It consists of dividing the studied area into grid cells and extracting time series of fire counts to fit the models. We use these models to estimate the fire season severity (FSS) defined as the sum of the fire counts detected in a season. 


After applying our approach to a global fire detection data sets, the results reveal that the forecasting errors are relatively low in the vast majority of the areas. These results indicate that it is possible to predict worldwide FSS by employing fire count historical data and time series forecasting methods. The proposed methodology can be especially useful for landscape managers and planners that lack the expertise and/or data for building or use some specialized forecasting model. It can also be applied to quickly estimate fire activity in large areas.


This manuscript is organized in the following form. We start by presenting in Section \ref{sec:matherial_and_methods} the data we use and the proposed forecasting methodology. In Sec. \ref{sec:results}, we present our results and discussion divided into three analyses: fire season length, fire season severity, and forecasting. Finally, we present in Sec. \ref{sec:conclusions} some conclusions and future works. 

\section{Material and Methods}
\label{sec:matherial_and_methods}

\subsection{Data}
\label{subsec:data}

In our experiments, we use data from the Moderate Resolution Imaging Spectroradiometer (MODIS) on-board NASA's Terra and Aqua satellites. Specifically, we used the Global Daily Fire Location Product (MCD14ML) Collection 6 from 2003 to 2017 \citep{dataset,giglio16,justice02}. This data set comprises active fire detection and is composed of geographic location, date, detection confidence, and some additional information for each fire pixel detected by the Terra and Aqua MODIS sensors. We consider only those fire records in the data set with detection confidence higher than 75\%. We disregard the years of 2001 and 2002 because they have missing data. The data set is freely accessible \citep{dataset}.

To analyze this spatiotemporal data set, we divide the globe into 65,612 hexagonal grid cells of approximated 7,774 $km^2$ each \citep{dggridR}. The main advantage of the hexagonal grid, when compared to the traditional rectangular longitude-latitude one, is that the former generates cells with more uniform coverage area and avoid distortions. Since the vast majority of the grid cells are located in regions without fire, like in the oceans or poles, we discard many cells. Only cells with at least one fire record per season were considered. After selecting only those cells with fire, we end up with 6486 ones. For each of those cells, we construct a time series with the daily fire records detected by the satellites. We use the term \emph{``fire counts''} (FC) to refer to these time series similarly as previous works \citep{chen11,earl18}.

Our choice for using fire counts instead of other types of fire data, such as burned area products, is simply justified by the characteristics of our analyses. This kind of data is easy to divide into grid cells and to interpret. Burned area information, on the other hand, would need to be reprocessed to be spatially reorganized into grid cells, adding complexity and not contributing with crucial information to our analyses. Another reason for using an active fire product is the availability of a daily timescale, required in our analysis. Furthermore, active fire data sets like the MCD14ML tend to detect smaller fires than burned area products \citep{earl17,fornacca17}, which leads to a satisfactory level of accuracy for our study. 

\subsection{Fire Season Estimation}
\label{subsec:fs_estimation}

\begin{figure}[!htb]
  \centering
  \includegraphics[width=0.78\linewidth]{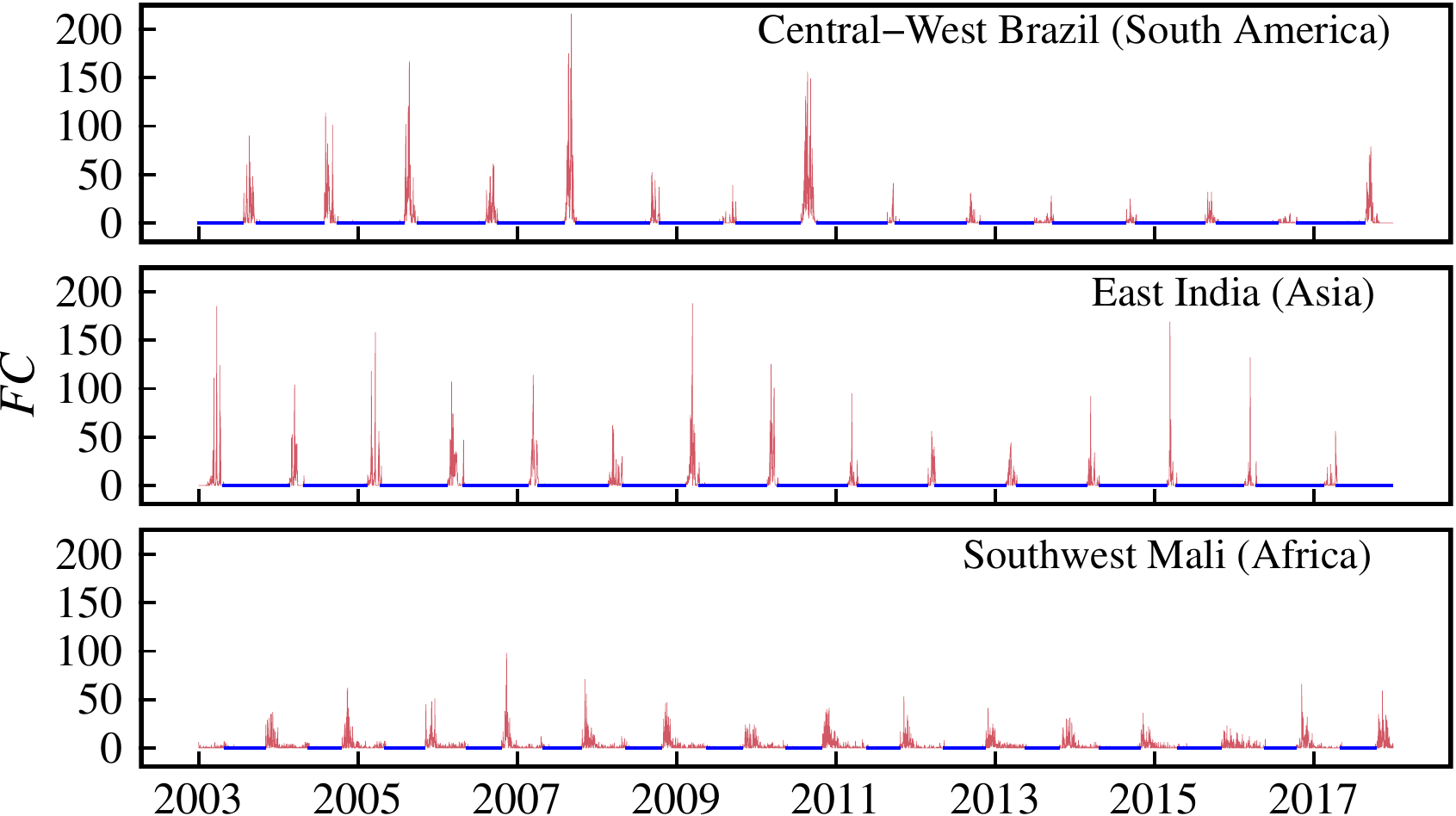}
  \caption{ Fire seasons estimation. Each time series represents the daily fire counts (FC) detected in grid cells located in three different regions: Brazil, India, and Mali. After the smoothing, we count the periods without fire (in blue) and consider the other periods (in red) as fire seasons.}
  \label{fig:season_detection}
\end{figure}

To estimate the length of the fire seasons, we propose a simple method: First, we apply a moving average (window size of seven days) to smooth the historical time series of fire counts. Then, we count the largest periods without fire. We remove these periods and consider the other ones as fire seasons. In Fig~\ref{fig:season_detection}, we illustrate this process using three FC time series from different regions. 

We define fire seasons as the time window centered in the month with the historical highest occurrence of fire in each cell \citep{chen11}. We use the fire season estimation method here proposed to define a single proper time window length to be used for all the cells. Specifically, this method was applied to the time series from all grid cells and verified the empirical distribution of fire season mean lengths. We remove outlier season lengths, i.e., values higher than the third quartile ($Q_3$) + 1.5 $\times$ interquartile range ($IQR$). Outlier lengths appear for multiple reasons. In some cases, these uncommonly long activities are accounted for false fire detections like in regions with hot bare soils \citep{Oom12}. In other cases, they represent volcanoes or gas flares from oil and gas exploration. The outlier removal minimizes the effect of these uncommon cells. After removing outliers, we choose the 99\% season length percentile as the global season length (see Sec.~\ref{subsec:fs_estimation}). This method guarantees that the FC time series from all cells have the same length and well represents the periods with more fire. The fire season severity (FSS) is the sum of fire counts (FC) in a season \citep{chen11}. 

\subsection{Forecasting methods and evaluation}
\label{subsec:forecasting_methods}

In our experiments, we discard the first and last season to avoid measuring broken seasons, resulting in 13 years of data. We also opt for using the monthly-accumulate fire counts (MA-FC) of each season to fit the forecasting models. The reason is the lack of FSS samples (13 in total), making it unfeasible to train the forecasting models. Since every season has seven months, the MA-FC has 91 values. We opt for using just the MA-FC in the seasons to train the models because the period outside the seasons is mainly formed by zeros (see Sec.\ref{sec:results}). 

\begin{figure}[!htb]
  \centering
  \includegraphics[width=0.78\linewidth]{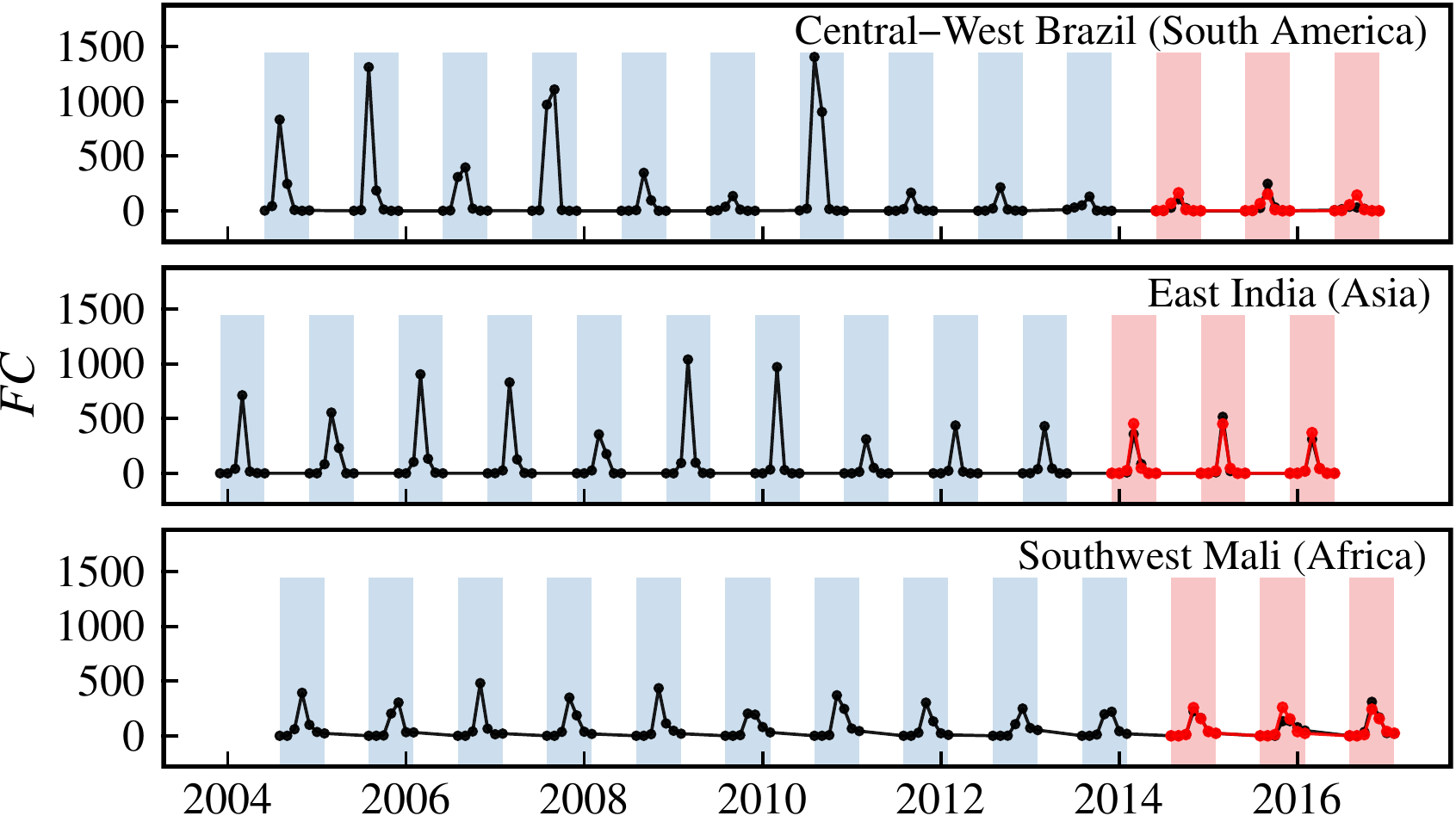}
  \caption{ Forecasting methodology. The three time series are the monthly-accumulate fire counts (MA-FC) for the examples in Fig.~\ref{fig:season_detection}. The first 10 seasons (blue background) are used to train the forecasting models and the last three ones (red background) are used to test. The time series in red are the result of the prediction using one of the methods (Prophet \citep{prophet17}). The final FSS forecast is the sum of the monthly forecasts from each season.} 
  \label{fig:fss_forecasting}
\end{figure}

For each cell, we separate the first ten seasons to train the model and the last three ones to test. We performed a Box-Cox transformation in both the train and test time series, and the parameter lambda was calculated using the method proposed by \citet{guerrero93}. The FSS forecast is the sum of the MA-FC predicted for a season. Our forecasting methodology is illustrated in Fig.~\ref{fig:fss_forecasting}, using three MA-FC time series (the same from Fig. \ref{fig:season_detection}) as examples with relatively high (Brazil), medium (India), and low (Mali) interannual variability for the peak months. In the three examples, the forecasting method (Prophet \citep{prophet17}) can reproduce the seasonal variations and accurately predict (in red) the test seasons.

\newpage
In the following, we briefly describe the forecasting methods:

\begin{itemize}
    
    \item \textit{Naive forecasting}: one of the simplest forecasting methods, it considers the last observation in a non-seasonal time series as the prediction for the next value \citep{hyndman18}. In a seasonal time series, the seasonal naive forecasting (\textit{snaive}) method considers each forecast to be equal to the observed value in the same period in the previous season. For example, the \textit{snaive} monthly prediction for July 2020 is the observed value in July of 2019. Although its simplicity, this method provides proper results in many cases. We use this method as a baseline.
    
    \item \textit{Autoregressive Integrated Moving Average models (ARIMA)}: aims to describe the autocorrelations on data given the past events, which is an approach widely employed for time series forecasting. The procedure starts analyzing the lag regression of the variable, and then, performing a moving average of errors from the linear combination of the past events. Therefore, the prediction of a variable is assumed to be a linear function of previous data and random errors. We employ the function in R that uses the Hyndman-Khandakar algorithm for automatic ARIMA modeling, which finds the best autoregression (AR) model and the moving average (MA) of weighted linear combination to obtain the prediction method~\citep{package_forecasting,liboschik17}.  

    \item \textit{Exponential smoothing (ETS)}: is a data forecasting method that uses a sliding window function to specify weights that decay exponentially over time. The sliding function helps for smoothing the series data, i.e., applying a low-pass filter to discard outliers or noise~\citep{hyndman18}. Thus, the previous data events are weighted with a geometrically decreasing ratio, i.e., the older the event, the lower its relevance weight. ETS methods can be used in datasets with seasonality, systematic trend, and other assumptions~\citep{liboschik17}. Besides, ETSs are considered a better option than the ARIMA methods because they are non-stationary and adopt exponentially decreasing weights calculated from previous events. Conversely, ARIMA methods are stationary and used linear weights of the past observations according to the moving average technique~\citep{hyndman18}.
    
    \item \textit{Short-term load forecasting (STLF)}: this approach can be defined as a time-series decomposition by seasons and trends using a Loess forecasting modeling~\citep{package_forecasting}. The method assumes that a time series can be separated in error, trend and seasonality components. Then, the results of the decomposition are used as input for the forecasting step. The time series is decomposed by a variation of a seasonal and trend function using Loess (STL)~\citep{cleveland1990STL}, which deals with multiple seasonality and nonlinear relationships. The returned multiple seasonal components are employed by a simple method for adjusting the seasonality of data~\citep{package_forecasting}, and the forecasting is the recombination process of the components. According to the authors, the method produces good forecasts results in seasonal time series~\citep{cleveland1990STL}.
    
    \item \textit{TBATS}: decomposes the time-series data into seasonal components and combines them in a trigonometric representation~\citep{Alysha2011}. These components are the Fourier terms of the series smoothed with an  exponential Box-Cox transformation~\citep{Alysha2011,package_prophet}. The trigonometric decomposition brings some advantages compared to traditional linear methods~\citep{Alysha2011}: it is possible to model non-integer seasonal data or high-frequency series, detecting more refined patterns; the method works with large parameter space not affecting the forecast results; it can process  nested or not seasonal components with nonlinear behavior, which is the case of real-world series data. Besides its capabilities, the method is robust to autocorrelation data, involving a simpler but efficient estimation procedure. 
    
    \item \textit{Generalized linear model for count time series (TSGLM)}: the method uses the generalized linear model (GLM) approach for modeling the data events (observations) according to previous conditional information~\citep{liboschik17}. For this purpose, it considers that the data are only composed of positive integers, i.e., count time series. In this way,  TSGLM provides a likelihood-based estimation capturing the dependence among data events~\citep{liboschik17}. This linear predictor provides a regression method over previous observations taking into consideration the covariance effects. For this point, it employs a probability distribution function, like Poisson (used in our experiments) or negative binomial, to estimate the maximum or quasi-likelihood of the distribution for obtaining the best statistical inference model~\citep{liboschik17}.
    
    \item \textit{Artificial neural networks (ANN)}: a model inspired by the biological neural networks constituted in animal brains. They are based on a set of connected units called artificial neurons \citep{bishop95}. The architecture of ANN frequently contains multilayer perceptron and sigmoid neurons, organized in layers and employing the stochastic gradient descent as the standard of the learning process. The ANNs are trained to learn the input-output relationships through an iterative process, in which the weight of the neurons are adjusted to minimize the error between the predicted and the true outputs. As a result, it is expected that the ANN learns a suitable model that accurately generalize the response to new data. In our experiments, we used Multi-Layer Perceptron (MLP) networks with a single hidden layer and lagged inputs \cite{package_forecasting}. We used just one seasonal lag as input and the hidden layer has half of the number of input nodes plus 1. So the number of neurons in the input, hidden, and output layers are respectively seven, five, and one. We repeat the training process for this topology 20 times starting with random weights initialization and we report only the best result.
    
    \item \textit{Prophet}: this is an open-source software released by Facebook~\citep{package_prophet} and used in several applications by the company for producing reliable forecasts in planning and goal setting. The algorithm follows an additive model approach where a non-linear smoother is applied to the regressor by yearly, weekly, and daily seasonality. The method analyzes three main components \citep{prophet17} trend, seasonality, and holidays effect, where the components are automatically detected from the data.  According to the authors, Prophet performs better than any other approach in most of the experimental results. They showed the robustness of the method in the presence of outliers, missing data,  and shifts in the trend~\citep{prophet17}. Besides, the method is suitable for time-series that present historical strong seasonal effects.
\end{itemize}

\newpage
In the experiments, we used the R programming language \citep{rstats} with the packages: \texttt{forecast} \citep{package_forecasting}, \texttt{prophet} \citep{package_prophet}, and \texttt{tscount} \citep{liboschik17}. These packages provide implementations of the forecasting methods and automatically tune the parameters for fitting the models to the time series. For details of implementation, we point the interested reader to the packages manuals and our source code described in Section \textit{``\nameref{sec:code}''}.

To find the best forecasting method of a single time series from a grid cell, we compared the mean absolute error (MAE) \citep{hyndman18}. Considering an observed (test) time series $Y$ of $n$ values and $F$ the predicted values for $Y$, the prediction error $e_t$ in a time $t$ is the difference between an observed value and its forecast: $e_t = Y_t - F_t$. The MAE is defined as:

\begin{equation}
    MAE = n^{-1} \sum\limits_{t=1}^{n}|e_t|.
    \label{eq:mae}
\end{equation}

The advantage of this measure is the easy interpretation of the error for a specific time series. However, the MAE is a scale-dependent measure that cannot be used to asses the accuracy in time series with different scales. The different scales make the accuracy interpretation not so clear in different scenarios (grid cells). To make the interpretation easier, in our experiments, we normalize the MAE by mean FSS values in each cell. The non-normalized results are presented in the \ref{appendix_forecasting_errors}.

We use the mean absolute scaled error (MASE) to compare the accuracies of different methods in all cells \citep{hyndman06}. The scaled forecasting error $q_t$ of a non-seasonal time series $Y$ is the error $e_t$ divided by the MAE of the non-seasonal naive forecast method on the training set:

\begin{equation}
    q_t = \frac{e_t}{\frac{1}{n-1}\sum\limits_{t=2}^{n}|Y_t-Y_{t-1}|}.
    \label{eq:mase_nonseasonal}
\end{equation}

The scaled error $q_t$ of a time series $Y$ with a seasonal period $m$ is:

\begin{equation}
    q_t = \frac{e_t}{\frac{1}{n-m}\sum\limits_{t=m+1}^{n}|Y_t-Y_{t-m}|}.
    \label{eq:mase_seasonal}
\end{equation}

\noindent If $q_t$ < 1, the forecasting method is better than the seasonal naive (\textit{snaive}), otherwise \textit{snaive} is better. As its name indicates, the MASE is simply the mean of all scaled errors:

\begin{equation}
    \label{eq:mase}
    MASE = n^{-1} \sum_{t=1}^{n}|q_t|,
\end{equation}

\noindent where $q_t$ corresponds to Eq. \ref{eq:mase_nonseasonal} if the modeled time series $Y$ is not seasonal or Eq. \ref{eq:mase_seasonal} otherwise. In our experiments, we use the MASE with Eq. \ref{eq:mase_seasonal} to evaluate the seasonal MA-FCs predictions. We use Eq. \ref{eq:mase_nonseasonal} to compare the non-seasonal FSS (sum of MA-FCs in a season) from the best model in each cell with the result of linear regression. We opt to compare to a linear regression due to the limited number of FSS values required to train other forecasting models. When MASE < 1, the forecasting method gives, on average, smaller errors than the one-step errors from the naive method. The best forecasting method has the lowest MASE values.

\section{Results and Discussion}
\label{sec:results}

\subsection{Fire season Analysis}

\begin{figure}[!htb]
  \centering
  \includegraphics[width=.95\linewidth]{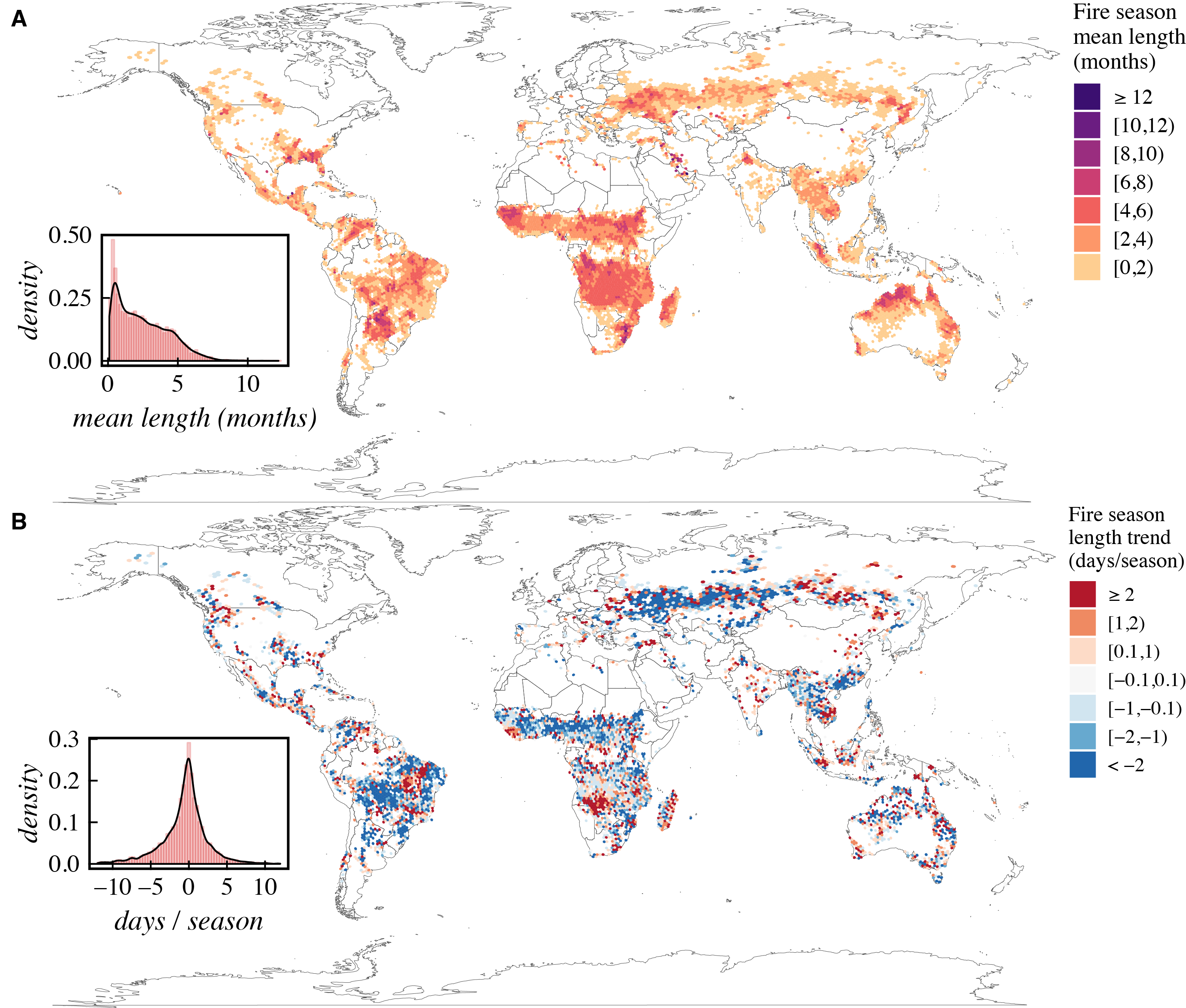}
  \caption{Global Fire season description: (A) length means and (B) linear trends (slopes). The inset figures show the respective probabilistic density functions (PDF).}
  \label{fig:fire_season_lengths}
\end{figure}

We propose in Sec.~\ref{subsec:fs_estimation}, a simple method to estimate the fire season length. In our first analysis, we apply this method to all the grid cells. Since we have 15 years of fire data, we extracted the same amount of fire seasons. In Fig. \ref{fig:fire_season_lengths}, we illustrate the fire season mean length and trend for each cell. Some of the regions with the longest (more than 7 months) fire season lengths are Paraguay, Southeast of the USA, north of Australia, and most parts of Africa. Since 99\% of the cells have fire seasons shorter than seven months (in the mean), we consider seven months a reasonable time window to represent worldwide fire seasons. More than half of the cells (57\%) present a decline in the fire season lengths but some regions like in Northeast Brazil, Eastern Russia, and parts of Africa show an increase in the fire season length. 

\begin{figure}[!htb]
  \centering
  \includegraphics[width=0.95\linewidth]{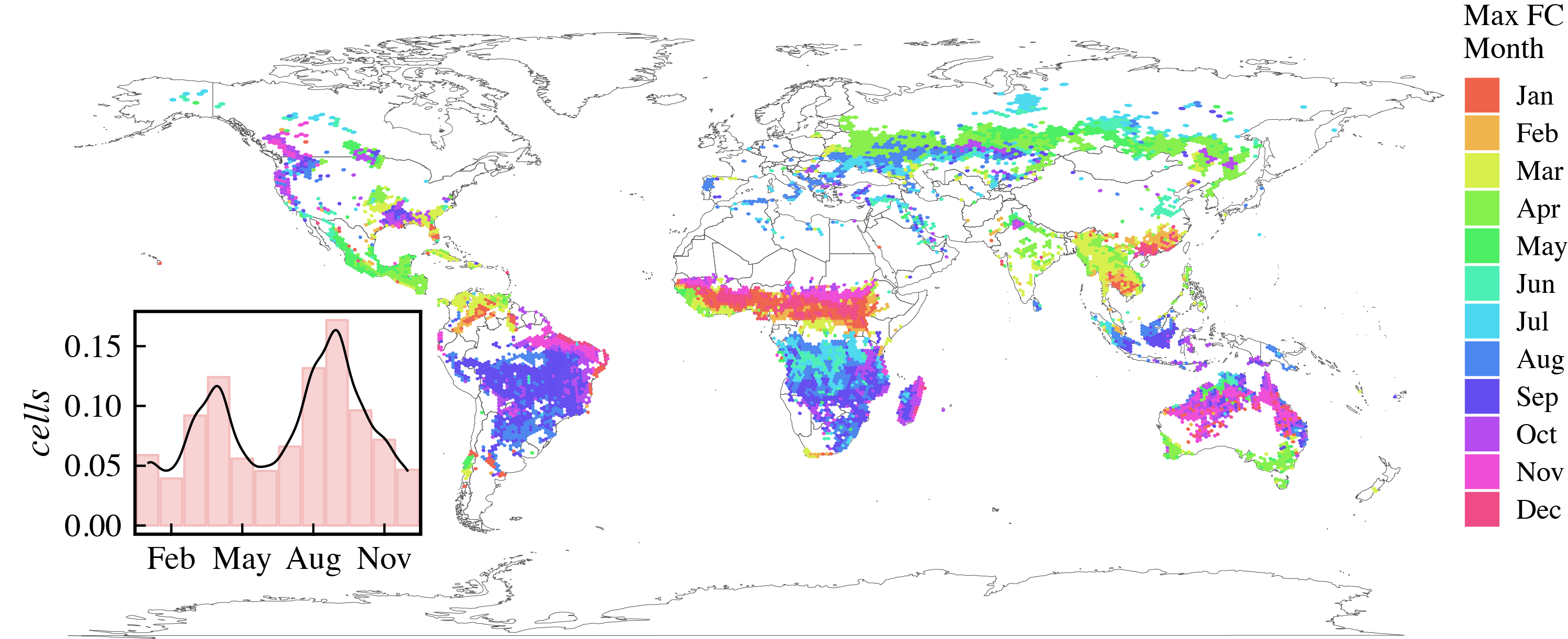}
  \caption{Months with the highest fire activity in each cell. The inset figure shows a histogram with the number of cells with the highest fire activity in a month.}
  \label{fig:fire_season_max_months}
\end{figure}

After defining seven months as the fire season length, we find the months with the historical highest occurrence of fire in each cell. Fig. \ref{fig:fire_season_max_months} illustrates the months with peaks of fire activity and a histogram with the months' distribution (inset). The months of historical maximum fire activity are distributed following a bimodal distribution (aligned with previous results \citep{benali17}). The two groups in the distribution mainly represent the dry seasons in the tropical rain belt (northern to the southern tropics). In the tropical zone, regions below the Equator have peaks of fire activity between August and October while locations above the Equator have maximums between March and May. Right on the equator, the fire activity is predominant at the end of the year (November to January). Part of the high fire activity in April is accounted for the dry seasons in parts of Russia and South Australia \citep{peel07}. 

\begin{figure}[!htb]
  \centering
  \includegraphics[width=0.95\linewidth]{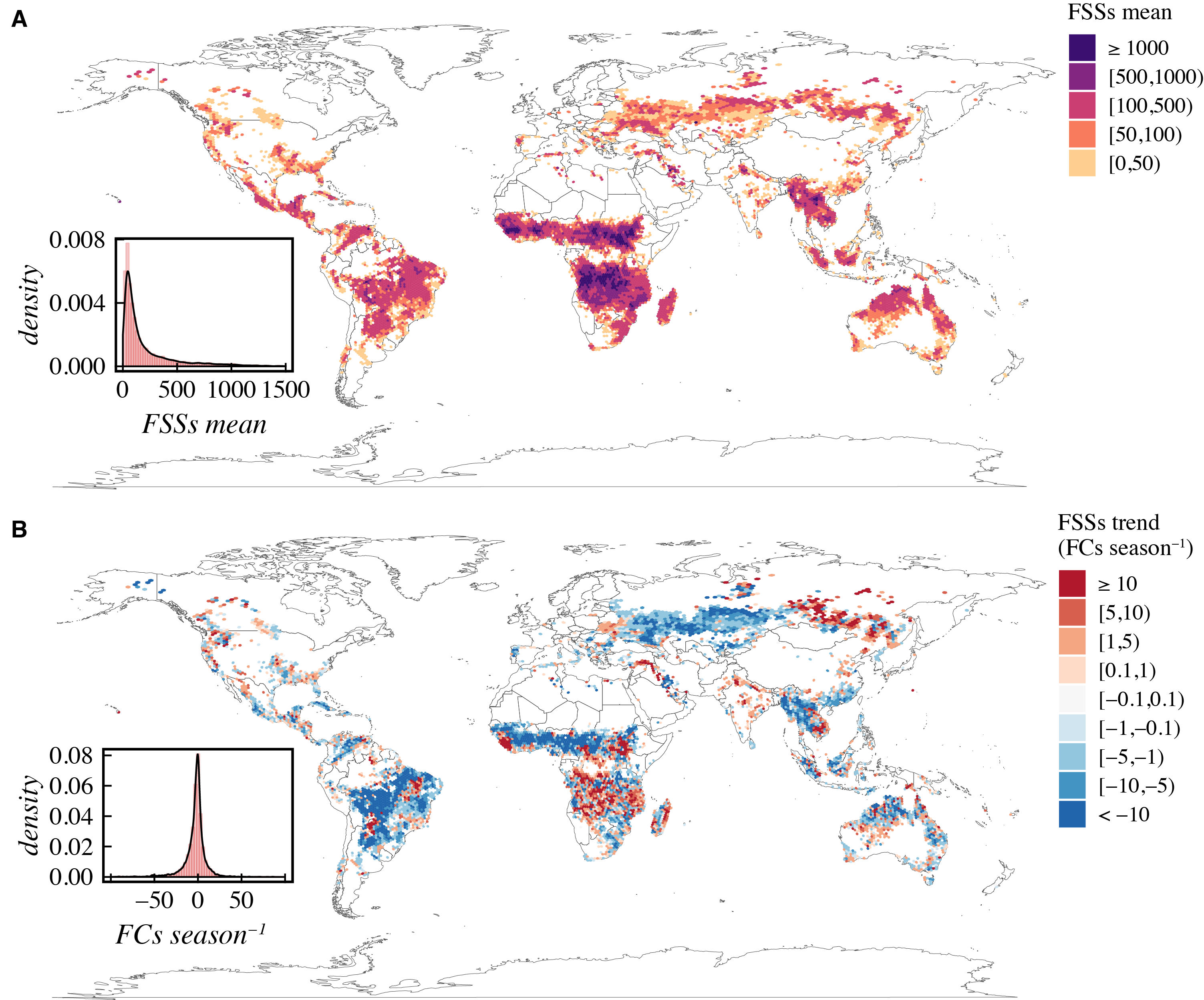}
  \caption{Global Fire season severity (FSS) description: (A) means (B) and linear trends (slopes). The inset figures show the respective probabilistic density functions (PDF). }
  \label{fig:fire_season_severity}
\end{figure}

The fire seasons are the periods of seven months centered in the month with the highest occurrence of fire.  In Fig.~\ref{fig:fire_season_severity}, we show the fire season severity (FSS) means and trends calculated for all the cells. 50\% of the global area has FSS mean lower than 100 FCs per year and 99\% have less than 1200 FCs on average. The highest number of fire counts concentrate in the tropics. In general, there is a trend of decline in the mean FSS worldwide (61\% of cells). However, in other regions like Southern Africa, Far Eastern Russia, Eastern Ukraine, and the West Coast of the USA, we can observe a notably increase tendency in the mean FSS.

Multiple reasons and factors can explain the trends in the fire seasons' lengths and severity. Some of the elements related to those variations are the increase or decrease of factors that lead to fire ignitions, such as less deforestation or lower lightning occurrence \citep{pereira08,bowman09,Oom12}. Climate phenomena, e.g. the El Niño-Southern Oscillation (ENSO), also influence the fire activity in critical regions like tropical forests \citep{pereira08}. Some previous works link the increase in global fire activity and seasonal length caused by climate change \citep{flannigan06,flannigan13,kelly13}. Similarly to our results, the global decrease in the fire activity was also observed in previous works that associate it to the worldwide increase in population densities and cropland areas \citep{andela17,arora18,earl18}.


\subsection{Fire season severity forecasting}

\begin{figure}[!htb]
  \centering
  \includegraphics[width=0.96\linewidth]{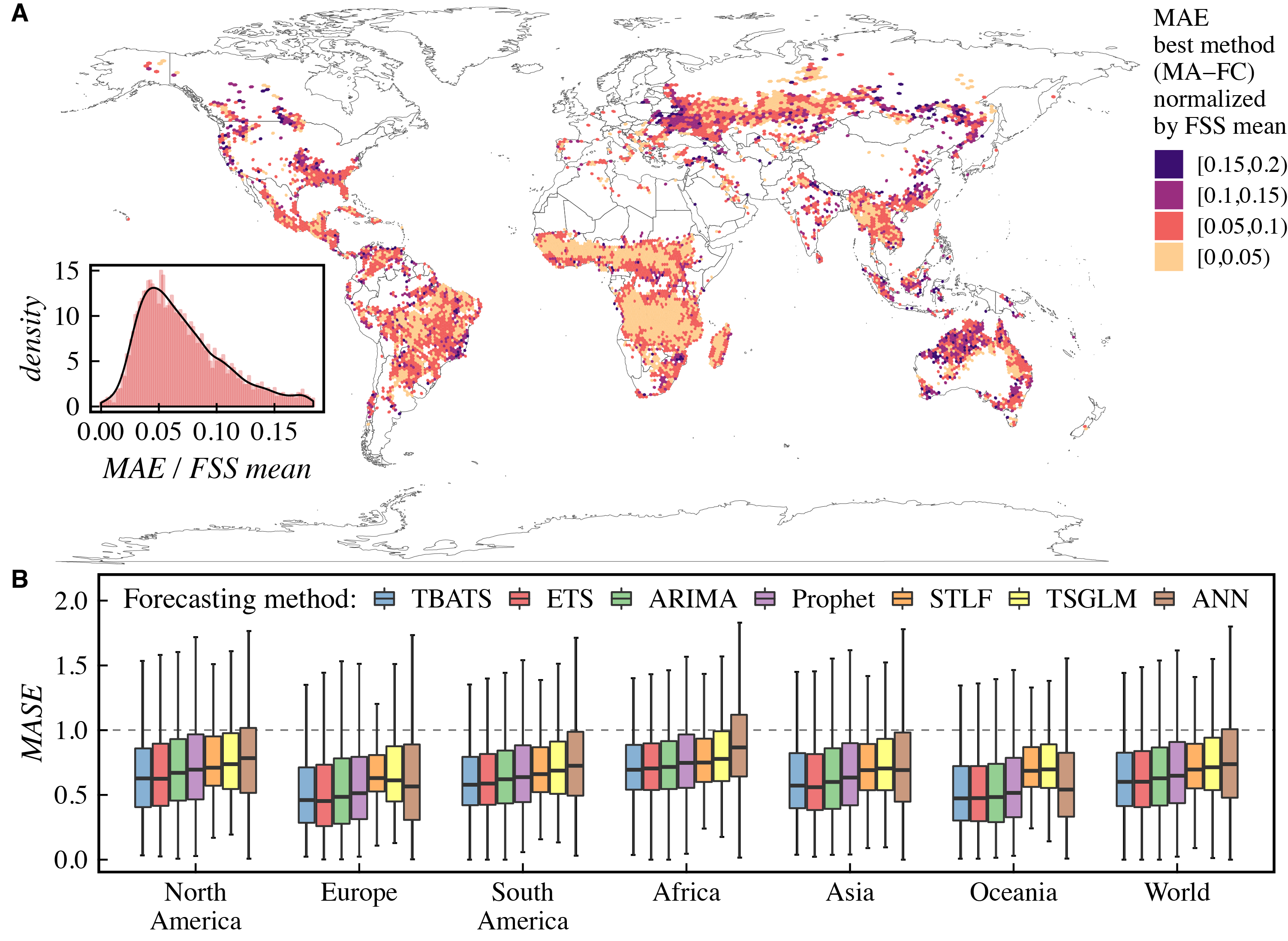}
  \caption{ Monthly-accumulated fire counts (MA-FC) forecasting. (A) The MAE from the best model for each cell normalized by the FSS mean (Fig \ref{fig:fire_season_severity}-A) and the PDF (inset). (B) Models comparison using the MASE (Eq. \ref{eq:mase_seasonal}) for the seasonal MA-FC from each cell divided by continent and the global. Outliers removed in both figures.}
  \label{fig:fire_season_forecasting}
\end{figure}

The forecasting results for the monthly-accumulated fire counts (MA-FC) are illustrated in Fig. \ref{fig:fire_season_forecasting}. On the top figure (\ref{fig:fire_season_forecasting}-A), we show the MAE for the best method normalized by the FSS mean (Fig \ref{fig:fire_season_severity}-A) and the density function for the MAE (inset). The regions with the higher relative MAEs are Northern Australia, Ukraine, parts of Russia, and the USA. Although, all the cells present MAE lower than the FSS mean, even in the African continent whose cells present that highest FSS means (Fig.~\ref{fig:fire_season_severity}). These results indicate a low general forecasting error. Since we defined FSS mean as a baseline, it is difficult to make general conclusions (especially in large regions) without taking into account the characteristics of each cell. Therefore, we point the reader interested in specific regions to the non-normalized MAEs, presented in the \ref{appendix_forecasting_errors} (Fig. \ref{fig:appendix1}).

In the bottom figure (\ref{fig:fire_season_forecasting}-B), we present a boxplot of the MASE for the entire world divided by continent. The TBATS and the ANN models present the lowest and the highest median error respectively. TBATS performs statistically better than the other methods except for the ETS ($p \leq$  0.001 -- Friedman and Nemenyi tests \citep{demsar06}). In 75\% of the cells, the forecasting methods provide results better (MASE < 1) than a seasonal naive forecasting method. Our results show that MAEs and MASEs for the MA-FC time series are low, which suggests that it is possible to apply historical detected active fire data to forecast fire seasons severities.

\begin{figure}[!htb]
  \centering
  \includegraphics[width=0.96\linewidth]{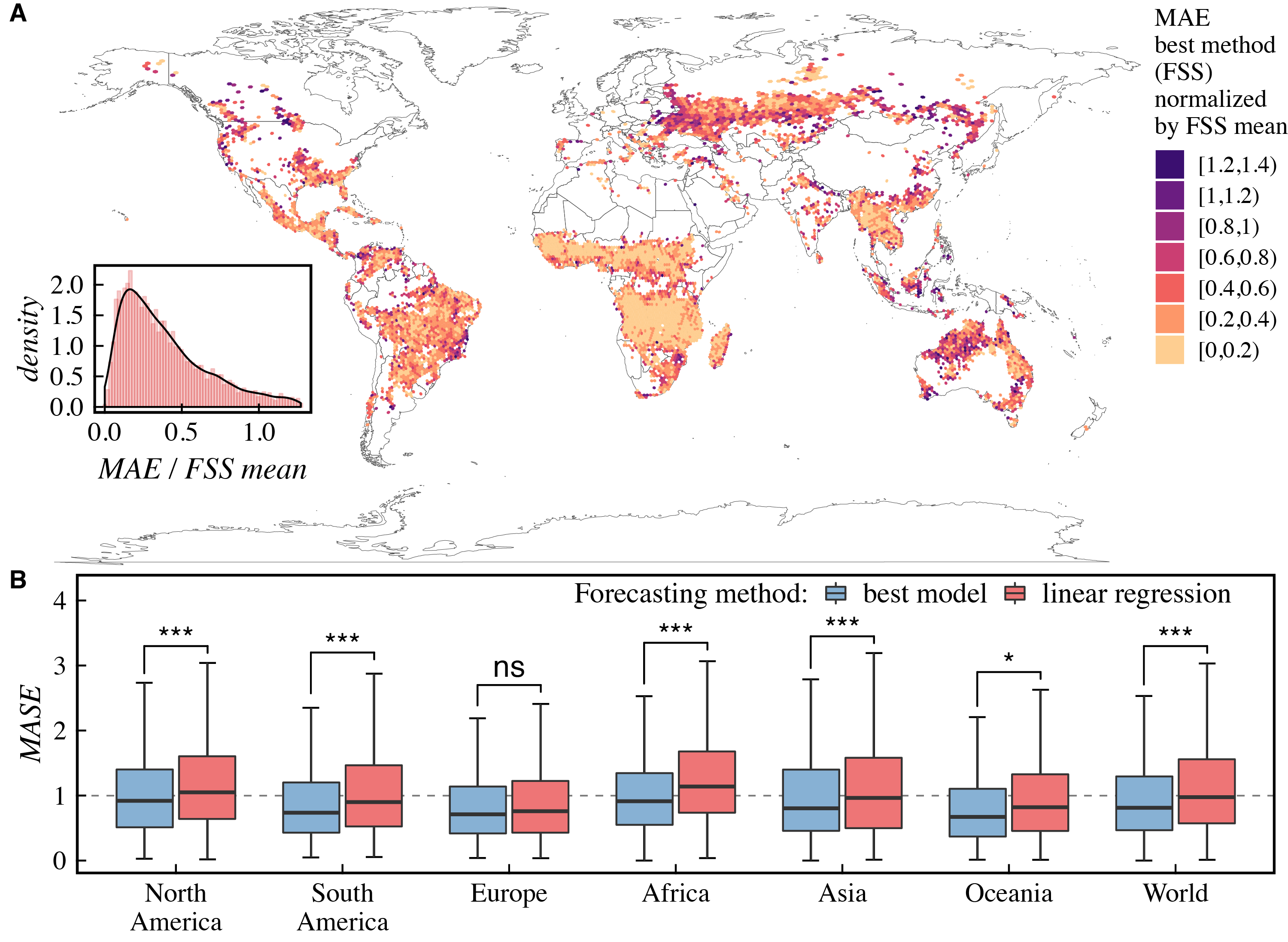}
  \caption{FSS forecasting. (A) The MAE from the best model for each cell and the PDF (inset). (B) Forecasting comparison between the best model achieved with MA-FCs and the linear regression in each cell. Accuracy was measured using the MASE (Eq. \ref{eq:mase_nonseasonal}) for the non-seasonal FSS from each cell divided by continent and the global (outliers removed). The significance thresholds are: not significant (\texttt{ns}), $p <$ 0.05 (*) and $p \leq$ 0.001 (***).}
  \label{fig:fire_season_forecasting_comparison}
\end{figure}

Given the MA-FC forecasting errors, we adopt the best model for each cell to make the FSS prediction. In Fig. \ref{fig:fire_season_forecasting_comparison}, we present the normalized MAEs and MASEs achieved comparing the best model for each cell with the FSS forecasting results from a linear regression. The non-normalized MAEs are reported in the \ref{appendix_forecasting_errors} (Fig. \ref{fig:appendix2}). In some regions, like north Australia and Ukraine, the normalized MAE is higher than 1, denoting that the errors are higher than the FSS mean values. Conversely, in most of the regions (95\%), the MAEs are lower than the FSS mean values. This result indicates worldwide predictability in the FSSs. The median MASEs are lower than one for the best models, which means that predictions are better than the naive forecasting method in at least 50\% of the cells (outliers removed). The best models for each cell statistically outperform the linear regression (according to the Wilcoxon paired test \citep{demsar06}), except in Europe where the errors are not statistically different. This result reinforces the worldwide predictability of FSSs in most of the regions.  

\section{Conclusions}
\label{sec:conclusions}

In this paper, we have analyzed global fire season severity, defined here as the accumulated fire counts detections in a season. We divided the globe into hexagonal grid cells and extracted time series of fire counts in each cell. We propose a very simple method to estimate the fire seasons. Our results show that 99\% of the cells have seasons shorter than seven months. We observed that the length of fire seasons are decreasing in general. However, in some regions like Southern Africa, Northeast Brazil, and Eastern Russia the fire seasons are increasing. Most of the cells (61\%) presented a decline in the FSS, excluding some regions like Southern Africa, Far Eastern Russia, Eastern Ukraine, and the West Coast of the USA. One possible explanation for the declining fire activity is the global agricultural expansion and intensification \citep{andela17}.

We have also shown that it is possible to forecast the FSS in many regions around the world. Since the FSS time series are very short, we used the monthly-accumulated fire counts (MA-FC) to train and test seven forecasting models. In this timescale, all the cells presented mean absolute error (MAE) lower than the FSS mean. The MAEs for the predicted FSS (sum of predicted MA-FC in a season) is lower than the FSS mean values in 95\% of the cells, indicating predictability in the global FSS. In general, the TBATS forecasting model provides the best results. We conclude that time series forecasting methods can be used to estimate worldwide fire seasons in a simple, fast and general way.

It is important to mention that our results are based on the historical data of active fire detections, which is short and limited. The historical data may not take into account climate change, which may considerably modify the fire seasons in the future. Our results are aligned with previous results \citep{andela17,arora18,earl18} but they might not reflect the future if the climate keeps changing. 

This work can be extended in many directions. Here we focused on global analysis but a natural next step is the thorough analyses of specific regions and the deep reasons that explain fire activity variations. Other variables, like socioeconomic projections and climate indices, can be included in the forecasting methods to improve the prediction. Other forecasting methods and machine learning tasks can also be explored on this data set. Tasks including fire seasons classification, clustering, and anomaly detection. These techniques might reveal new patterns about the interplay between anthropogenic activity and global fire dynamics. 

\section*{Acknowledgments}

This research is supported by the Fundação de Amparo à Pesquisa do Estado de São Paulo (FAPESP) under Grant 2015/50122-0 and the German Research Council (DFG-GRTK) Grant 1740/2. L.N.F. acknowledges FAPESP Grants 2019/00157-3 and 2017/05831-9. D.A.V.O acknowledges FAPESP Grants 2016/23698-1, 2018/01722-3, and 2018/24260-5. E.E.N.M acknowledges FAPESP Grant 2018/03211-6. This research was developed using computational resources from the Center for Mathematical Sciences Applied to Industry (CeMEAI) funded by FAPESP (grant 2013/07375-0). 

\section*{Computer Code Availability}
\label{sec:code}

The fire season estimation method and the forecasting algorithms used in this paper were implemented using the R programming language \citep{rstats} and the packages: \texttt{forecast} \citep{package_forecasting}, \texttt{prophet} \citep{package_prophet}, and \texttt{tscount} \citep{liboschik17}. The R software, packages, and our implementation are open-source and freely available for download. All the packages are distributed under GPL license, except \texttt{prophet} that is distributed under BSD-3-Clause license. Our source code can be downloaded in the following link:

\vspace{0.5cm}

\centerline{\texttt{\href{https://lnferreira.github.io/global\_fss\_analysis\_forecasting/}{https://lnferreira.github.io/global\_fss\_analysis\_forecasting/}}}

\newpage
\appendix

\section{Forecasting Errors}
\label{appendix_forecasting_errors}

\begin{figure}[!htb]
  \centering
  \includegraphics[width=1\linewidth]{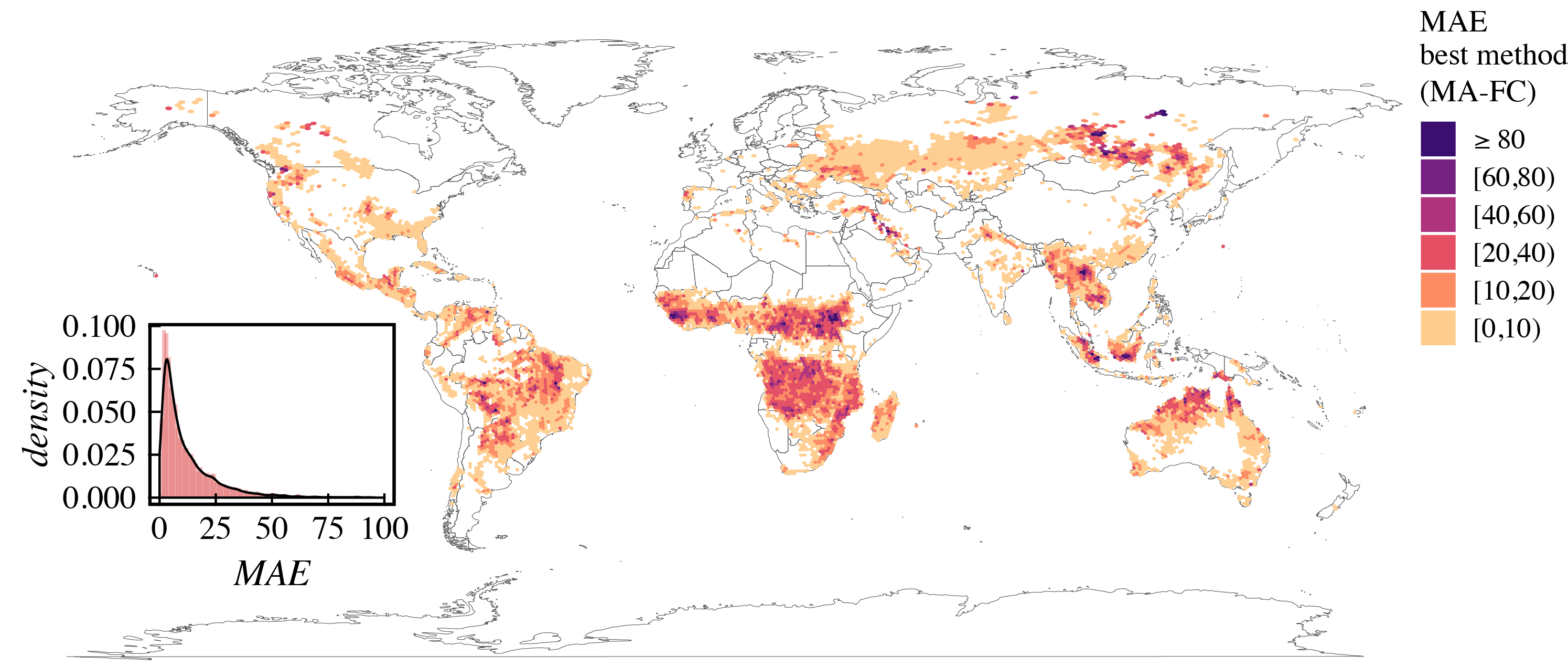}
  \caption{ Monthly-accumulated fire counts (MA-FC) forecasting. The MAE from the best model for each cell and the PDF (inset).}
  \label{fig:appendix1}
\end{figure}

\begin{figure}[!htb]
  \centering
  \includegraphics[width=1\linewidth]{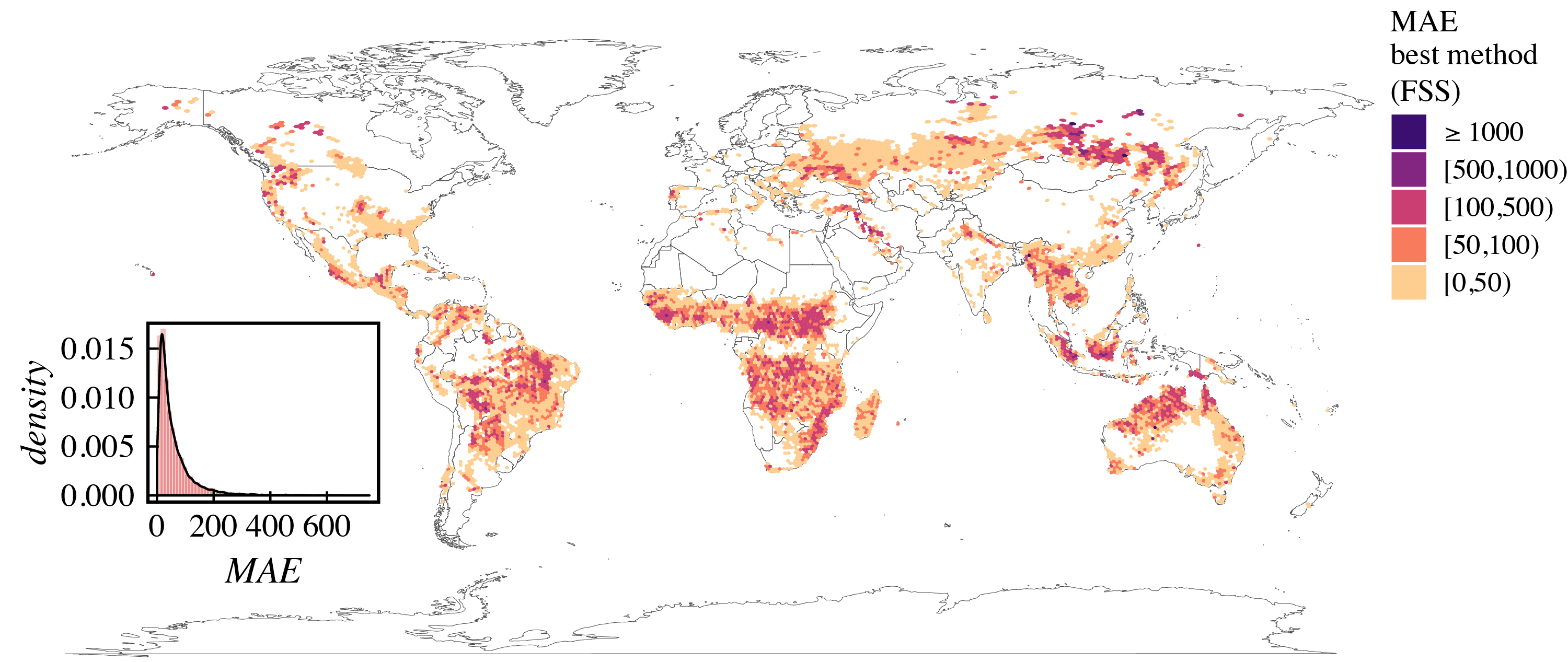}
  \caption{ FSS forecasting. The MAE from the best model for each cell and the PDF (inset).}
  \label{fig:appendix2}
\end{figure}


\newpage




\bibliographystyle{model2-names.bst}\biboptions{authoryear}





\bibliography{references.bib}

\end{document}